\documentclass[aip,pop,floatfix,
reprint,
groupedaddress]{revtex4-1}
\usepackage{amsmath}
\usepackage{amssymb}
\usepackage{graphicx}%
\usepackage{verbatim}
\renewcommand{\vec}[1]{\mathbf{#1}}
\begin{document}

\title{Generation of Electrojets in Weakly Ionized Plasmas through a Collisional
Dynamo}
\author{Y. S. Dimant}
\email[]{dimant@bu.edu}
\author{M. M. Oppenheim}
\author{A. C. Fletcher}

\affiliation{Center for Space Physics, Boston University, 725 Commonwealth Ave., Boston, MA 02215}
\date{\today}

\begin{abstract}
Intense electric currents called electrojets occur in weakly ionized
magnetized plasmas.  An example occurs in
the Earth's ionosphere near the magnetic equator where neutral winds
drive the plasma across the geomagnetic field. Similar processes take
place in the Solar chromosphere and MHD generators. This letter argues that
not all convective neutral flows generate electrojets and it introduces the
corresponding universal criterion for electrojet formation, $\nabla\times
(\vec{U}\times\vec{B})\neq\partial\vec{B}/\partial t$, where $\vec{U}$ is the
neutral flow velocity, $\vec{B}$ is the magnetic field, and $t$ is time. This
criterion does not depend on the conductivity tensor,
$\hat{\sigma}$. For many systems, the displacement current,
$\partial\vec{B}/\partial t$, is negligible, making the criterion even
simpler. This theory also shows that the neutral-dynamo driver that generates
electrojets plays the same role as the DC electric current plays for the
generation of the magnetic field in the Biot-Savart law.
\end{abstract}


\maketitle


A weakly ionized plasma in a strong magnetic field, colliding with a
neutral gas, can generate electric currents called electrojets. One
such neutral-driven electrojet, named the equatorial electrojet,
forms in the Earth's E-region ionosphere around the magnetic equator
\cite{Rishbeth:Ionospheric97, Forbes:Equatorial81,
Heelis:Ionosphere04, Kelley:Ionosphere2009}. This electrojet results from a large
$\vec E$-field that ultimately derives its energy from neutral winds,
abundant in the bottom of the thermosphere (90 - 130 km altitude).
The resulting $\vec{E} \times\vec{B}$ drifting electrons cause the
primary electrojet current.  Similar electrojets exist along the magnetic
equators of other magnetized planets \cite{Raghavarao:Dynamo83}.
Strong convective neutral flows across $\vec B$ in the highly
collisional solar chromosphere will also generate electrojets
\cite{Fontenla:Chromospheric05,Fontenla:Chromospheric08,Madsen:Multi14}. Under
special conditions, similar processes can take place in
magnetohydrodynamic (MHD) generators
\cite{Rosa:Plasmas91,Bokil_Toward15}.  This paper presents a novel
approach in analyzing the strength of electrojets and develops a
universal criterion for the existence of these currents. It also
shows that the simple 1D model often used by textbooks and papers to
illustrate the origin of electrojets will not, in fact, create an
electrojet.

Electrojets develop a complex array of behaviors beyond just
generating currents strong enough to cause large magnetic field
perturbations.  These currents also frequently drive various plasma
instabilities that result in plasma density irregularities and
fluctuating electric fields \cite{Farley:Equatorial09}.
These irregularities and fields have been
observed for a long time by radars and rockets
\cite{Kelley:Ionosphere2009,Pfaff:Near-Earth12}.  These instabilities
can cause intense electron heating and anomalous conductivities
\cite{Schlegel:Anomalous81,Foster:Simultaneous00,DimantMilikh:JGR03,Dimant:Magnetosphere2011}.
\citet{Fontenla:Chromospheric05} has speculated that such
instabilities may play an important role in chromospheric heating.

Theoretical studies of the Earth's equatorial electrojet have a long
history \cite{Bramley:Effects67, Krylov:Structure74,
Richmond:Equatorial73-I, Forbes:Atmospheric76_II}, providing
detailed quantitative descriptions of the electrodynamic effects due
to different components of the neutral wind.  However, they have never
addressed the following simple questions: 1) Do the neutral convection
flows always drive electrojets? and 2) What component of the winds
drive electrojet formation? We answer these fundamental questions in
this letter. We provide a simple universal criterion for wind-driven
electrojets and identify the driving components of the wind. We
demonstrate that this driver plays the same role in electrojet
formation as does the DC electric current in the generation of
magnetic field via the Biot-Savart law.

Auroral electrojets, found at high latitudes of magnetized planets,
result from externally imposed electric fields that propagate along
field lines from well outside the electrojet region. The neutral wind
has only modest effects on these electrojets and this analysis does
not apply.


Electrojets form in plasmas where the neutral density is sufficient to
collisionally demagnetize the ions, $\Omega_{i}\ll\nu_{i}$, but not
the electrons, $\Omega_{e}\gg\nu_{e}$, where $\Omega_{e,i}$ are the
electron/ion gyrofrequencies and $\nu_{e,i}$ are the
electron-neutral/ion-neutral collision frequencies. In a spatially
inhomogeneous plasma, the convective neutral flow affects each species
differently, resulting in a small charge separation. This generates an
ambipolar electrostatic field that leads to the formation of
electrojets, provided the conditions derived herein are satisfied.

This analysis focuses on large-scale and slow evolution, and so assumes a weakly
ionized, inertialess, cold, and quasineutral plasmas. This
leads to the following dynamo equations,
\begin{equation}
\nabla\cdot\vec{J}=0,\qquad\vec{J}=\hat{\sigma}(\vec{E}+\vec{U}\times\vec{B}),
\label{initial}%
\end{equation}
where $\vec{J}\equiv\sum_{\alpha}q_{\alpha}n_{\alpha}\vec{V}_{\alpha}$ is the
total plasma current density, $q_{\alpha}$, $n_{\alpha}$, and $\vec{V}%
_{\alpha}$ are respectively the charge, particle density, and mean fluid
velocity of species $\alpha$, which includes including multiple ion species and
electrons. In Eq.~(\ref{initial}), $\hat{\sigma}$ is the anisotropic conductivity tensor
determined in the local frame of reference of the neutral flow, $\vec{E}$ and $\vec{B}$ are
the total DC electric and magnetic fields, and $\vec{U}$ is the convective
velocity of the neutral gas \cite{Kelley:Ionosphere2009}.

Now we will obtain a simple criterion for electrojet formation. The
$\vec{U}\times\vec{B}$ term in eq.~(\ref{initial}) underlies
neutral-dynamo driven electrojets.
The neutral flow and magnetic field are usually decoupled, so that
$\vec{U} \times\vec{B}$ can form a general vector field. In the
simplest case, one can assume that the magnetic field is stationary,
$\partial\vec{B}/\partial t=0$, and the dynamo term is irrotational,
$\nabla\times(\vec{U}\times\vec{B})=0$, so that
$\vec{U}\times\vec{B}=-\nabla\Psi$.   In this case,
Eq.~(\ref{initial}) becomes $\vec \nabla \cdot \hat{\sigma}(\vec \nabla \Phi
+\vec \nabla \Psi)=0$, where $\Phi$ is the electrostatic potential.  One
solution exists when $\Phi=-\Psi$.  This is a unique stable solution when
$\Phi=-\Psi =0$ on the boundaries.  A proof of this is beyond the
scope and scale of this paper.

This means that, for a plasma embedded in a dense neutral flow with a
large collisional momentum exchange, the frictional forces reduce the
differences between the convection velocities of the neutral gas and
plasma. The resulting $\vec{E}$-field creates a quasi-neutral plasma
flow that satisfies $\vec{E}+\vec{U}\times\vec{B}=0$.  Physically,
this means that there is no difference between the mean fluid
velocities, $\vec{V}_{e}=\vec{V}_{i}=\vec{U}$, resulting in
$\vec{J}=0$. This state is achieved for an arbitrary conductivity
tensor, $\hat{\sigma}$. For magnetized electrons and unmagnetized ions, such a plasma becomes `frozen' into the
neutral flow during a short relaxation time,
$\tau_{\mathrm{rel}}\sim\max[\Omega_{e}^{2}/(\nu_{e}\omega_{pe}^{2}),\
\nu_{i} ^{-1}]$,
where $\omega_{pe}$ is the electron plasma frequency. In this frozen
flow, the motion of magnetized electrons is sustained by the
$\vec{E}\times\vec{B}$ drift, while the ion motion is sustained mostly
by ion-neutral collisions.

For the case when
\begin{equation}
\nabla\times(\vec{U}\times\vec{B})\neq0,\label{criterion}%
\end{equation}
no charge separation can create an electric potential that
perfectly matches the neutral drag and, hence, completely cancels the
current. Therefore, Eq.~(\ref{criterion}) represents the criterion for
driving an electrojet by a neutral dynamo with stationary $\vec{B}$. This
criterion is valid for an arbitrary conductivity tensor and has to be fulfilled at
least somewhere within the conducting plasma.

The Eq.~(\ref{criterion}) criterion is easily extended to
non-stationary $\vec{B}$. Separating the total electric field
$\vec{E}$ into
inductive, $\vec{E}_{\mathrm{ind}}$, and electrostatic, $-\nabla
\Phi$, parts,  and using Faraday's equation, we obtain the inequality
\begin{equation}
\nabla\times(\vec{U}\times\vec{B})\neq\frac{\partial\vec{B}}{\partial t},
\label{irrotational}
\end{equation}
that gives the general condition for a neutral dynamo to drive electrojets.

This general criterion stays the same even after
including the plasma pressure and gravity -- factors neglected in the standard
dynamo described by Eq.~(\ref{initial}).
A weakly ionized plasma behaves as an isothermal gas, such that the
pressure terms in the fluid momentum equations, $-\nabla P_{\alpha
}/n_{\alpha}$, can be expressed as $-\nabla\left(  T_{\alpha}\ln n_{\alpha}\right)$.
Since the gravity force is always the gradient of a gravitational
potential, both the pressure terms and gravity can be expressed as gradients
of scalar functions which can be combined with $\Phi$ with no consequences for
Eq.~(\ref{irrotational}).

In order to better understand the criterion expressed by
Eq.~(\ref{irrotational}), we expand its left-hand side in a standard way and
apply the continuity equation for the neutral flow, $\partial\rho/\partial
t+\nabla\cdot(\rho\vec{U})=0$, where $\rho$ is the neutral gas density.
Introducing the convective derivative, $D/Dt\equiv\partial/\partial_{t}%
+\vec{U}\cdot\nabla$, Eq.~(\ref{irrotational}) becomes
\begin{equation}
\frac{D}{Dt}\left(  \frac{\vec{B}}{\rho}\right)  \neq\left(  \frac{\vec{B}%
}{\rho}\cdot\nabla\right)  \vec{U}.\label{generalized_sootno}%
\end{equation}
This inequality means that the magnetic field is not `frozen' into the neutral
flow.
Many simple systems will not meet this requirement.  For example, an incompressible 2D
neutral flow perpendicular to a constant $\vec{B}$ cannot generate an
electrojet, regardless of the spatial inhomogeneity of the conductivity
tensor.


Though Eq.~\ref{irrotational} tells us that the curl components of $\vec{U}
\times \vec{B} + \vec{E}_{\mathrm{ind}}$ cause electrojets, it does not extract those
components.  We will now do so. Defining
$\vec{K}\equiv\vec{U}\times\vec{B}+\vec{E}_{\mathrm{ind}}$, we simplify
Eq.~(\ref{irrotational}) to $\nabla\times\vec{K}\neq0$.
The Helmholtz decomposition of $\vec{K}$, assuming that $\vec{K}(\vec{r},t)$
vanishes at infinity sufficiently rapidly \cite{Arfken:Mathematical05},
gives
\begin{equation}
\vec{K}(\vec{r},t)=-\ \nabla\Psi+\nabla\times\vec{A},\label{Y=}%
\end{equation}
where the scalar and vector `potentials' are given by 3D volume integrals%
\begin{subequations}
\begin{equation}
\Psi(\vec{r},t)=\frac{1}{4\pi}\int\frac{\nabla^{\prime}\cdot\vec{K}(\vec{r}^{\prime}%
,t)}{|\vec{r}-\vec{r}^{\prime}|}\ d^{3}r^{\prime},\label{Psi,}
\end{equation}
\begin{equation}
\vec{A}(\vec{r},t)=\frac{1}{4\pi
}\int\frac{\nabla^{\prime}\times\vec{K}(\vec{r}^{\prime},t)}{|\vec{r}-\vec
{r}^{\prime}|}\ d^{3}r^{\prime}.\label{A}%
\end{equation}
\label{Psi,A}
\end{subequations}
Here $\nabla^{\prime}$ indicates that the corresponding vector
differentiations are with respect to $\vec{r}^{\prime}$. In 2D problems,
$d^{3}r^{\prime}/|\vec{r}-\vec{r}^{\prime}|$ should be replaced by $-\ln(|\vec
{r}-\vec{r}^{\prime}|^{2})d^{2}r^{\prime}$.
If we consider a finite volume restricted by the boundary surface $S$, then
Eq.~(\ref{Y=}) holds with slightly modified `potentials'. The 3D integrations
in Eq.~(\ref{Psi,A}) outside the finite volume are replaced by 2D integrals
over the boundary surface with $\nabla^{\prime}$ replaced by $-\hat{n}$, where
$\hat{n}(\vec{r}^{\prime})$ is the local unit normal to the surface $S$,
directed outward.

In the expression for the total current density, $\vec{J}=\hat{\sigma}%
(-\nabla\Phi+\vec{K})$, the scalar `potential' $\Psi$ can be combined with the
actual electrostatic potential, $\Phi$, into one scalar function, $\Upsilon\equiv\Phi+\Psi
$, so that the current density becomes
\begin{equation}
\vec{J}=\hat{\sigma}\left(  -\nabla\Upsilon+\nabla\times
\frac{1}{4\pi}%
\int\frac{\nabla^{\prime}\times\vec{K}(\vec{r}^{\prime},t)}{|\vec{r}-\vec
{r}^{\prime}|}\ d^{3}r^{\prime}
\right)  , \label{J=}%
\end{equation}
where
\begin{subequations}
\label{rotor}%
\begin{align}
\nabla^{\prime}\times\vec{K}(\vec{r}^{\prime},t)  &  =\nabla^{\prime}%
\times(\vec{U}\times\vec{B})-\frac{\partial\vec{B}}{\partial t}%
\label{rotor_Y_1}\\
&  =\rho\left[  \frac{D^{\prime}}{Dt}\left(  \frac{\vec{B}}{\rho}\right)
-\left(  \frac{\vec{B}}{\rho}\cdot\nabla^{\prime}\right)  \vec{U}\right]  .
\label{rotor_Y}%
\end{align}

In the general case, 
$\vec{J}$ is not divergence-free, so
that the quasi-neutrality equation $\nabla\cdot\vec{J}=0$ leads to a
second-order partial differential equation for the unknown scalar function
$\Upsilon$,
\end{subequations}
\begin{equation}
\nabla\cdot\left(  \hat{\sigma}\nabla\Upsilon\right)  =\nabla\cdot\lbrack
\hat{\sigma}(\nabla\times\vec{A})].\label{urka}%
\end{equation}
Equation~(\ref{urka}) is equivalent to Eq.~(\ref{initial}), except that it shows
explicitly that $\nabla\times\vec{K} =
\nabla\times(\vec{U}\times\vec{B})-\partial \vec{B} / \partial t
$ plays, for electrojet generation, the same
role as DC electric currents play in the Biot-Savart law for magnetic
field generation.
The gradient of the scalar function $\Upsilon$ is the
field that forms electrojets.

In the simplest case of spatial uniformity and an
isotropic and unmagnetized conductivity tensor of  $\sigma_{ij}=\sigma\delta_{ij}$,
the current density $\vec{J}$ in Eq.~(\ref{J=}) becomes divergence-free with
$\Upsilon=0$. When $\nabla\times\vec{K}\neq0$ a current develops with $\vec
{J}=\nabla\times(-\sigma\vec{A}\mathcal{)}$, where $-\sigma\vec{A}$ is fully
equivalent to the volume integral in the Biot-Savart law.

In the general case of an anisotropic and spatially inhomogeneous conductivity
tensor $\hat{\sigma}$, the electrojet generation is more complicated than the
magnetic field generation described by the Biot-Savart law. The vector field
$\nabla\times\vec{K}$, with the corresponding integration in $\vec{A}$,
represents only the primary source, while the entire electrojet formation
undergoes an additional step.
For the anisotropic and non-uniform $\hat{\sigma}$, we have $\nabla\cdot
\lbrack\hat{\sigma}(\nabla\times\vec{A})]\neq0$ in almost all locations. In
these locations, the unhindered current $\hat{\sigma}(\nabla
\times\vec{A})$ would accumulate significant volume charges. As soon
as this accumulation starts, a non-uniform electrostatic potential
$\Phi=\Upsilon-\Psi$ forms, making the entire current, $\vec{J}
=\hat{\sigma}\nabla(\Upsilon-\nabla \times\vec{A})$,
divergence-free and, hence, preventing further charge
accumulation. Finding the sustained spatial distribution of the
quasi-stationary potential $\Phi$, and hence of the total current
density $\vec{J}$, requires solving Eq.~(\ref{initial}).

Unlike Eq.~(\ref{initial}), however, Eq.~(\ref{urka}) allows us to
identify $\nabla\times\vec{K}$
as the actual driver of electrojet formation.  Equation~(\ref{urka})
provides a smooth transition to $\nabla\times \vec{K} = 0$ when the `vector-potential'
$\vec{A}$, and hence the right-hand side of Eq.~(\ref{urka}), goes
away, resulting in $\Upsilon = 0$ and no electrojet. Similar to the
electric current in the Biot-Savart law, the electrojet driver
$\nabla\times\vec{K}$ does not have to be non-zero everywhere. For
example, imagine a situation where a closed neutral flow,
confined within a small volume, generates an electrojet that occupies a much
larger space, like a large-scale magnetic field generated by a
localized magnetic dipole.


The criterion for electrojet formation should arise directly from explicit
analytical solutions of Eq.~(\ref{initial}) or (\ref{urka}). To trace this, we consider three
simplified models of electrojets that allow such solutions.
These models exhibit some key features of the actual electrojets.

The first model is a generalization of the trivial model presented in
some textbooks and review papers \cite{Kelley:Ionosphere2009,Farley:Equatorial09,Pfaff:Near-Earth12},
and shows that this model doesn't actually yield electrojets.  The second is an axially symmetric model with a geometry
similar to MHD generators.  The third invokes a more complex
magnetic field and is a simplified 2-D approximation of the equatorial electrojet
and the solar chromosphere.


The first model is the simplest case that produces an electrojet
though, as we will see, not directly at the magnetic equator. We assume a
horizontally stratified neutral flow, embedded in a uniform magnetic field
$\vec{B}=B\cos I\,\hat{e}_{x} + B\sin I\,\hat{e}_{z}$ and
$\vec{U}=U_{x}(z)\hat{e}_{x}+U_{y}(z)\hat{e}_{y}$.    The constant magnetic field  $\vec{B}$
points in an arbitrary direction with respect to
the horizontal plane (the inclination, or `dip' angle),
$I$.  We require $U_{x,y}(z) \rightarrow 0$ as $z \rightarrow 0,\infty$, while
$\vec{B}$, $\rho(z)$ and $n(z)$ occupy the entire space.
As shown below, the plasma
will also form a horizontally stratified flow, so that the vertical
profiles of the neutral gas and plasma densities, $\rho(z)$ and $n(z)$, are
unaffected by the flows and can be independently specified.

The 1D quasi-neutral equation $\partial
J_{z}/\partial z=0$ yields a constant $z$-component of the current density,
$J_{z}=J_{z0}$. Assuming no imposed external currents, we set $J_{z}=0$.
In a 1D system with no potential on the boundaries, the electric
field can only have a vertical component, $E_{z}=-\partial\Phi/\partial z$.
Using this information, $\vec{J}=\hat{\sigma}(\vec{E}+\vec
{U}\times\vec{B})$ becomes
\begin{widetext}
\begin{equation}
J_{x} = \frac{\sigma_{\parallel}\left(  \sigma_{\mathrm{H}}U_{x}\sin
I+\sigma_{\mathrm{P}}U_{y}\right)  }{\sigma_{\parallel}\sin^{2}I+\sigma
_{\mathrm{P}}\cos^{2}I}\, B_{z},\qquad
J_{y} = \frac{\sigma_{\parallel}\left(  \sigma_{\mathrm{H}}U_{y}%
-\sigma_{\mathrm{P}}U_{x}\sin I\right)  \sin I-\left(  \sigma_{H}^{2}
+\sigma_{P}^{2}\right)\!  U_{x}\cos^{2}\!I}{\sigma_{\parallel}\sin^{2}
I+\sigma_{\mathrm{P}}\cos^{2}I}\, B_{z},\label{J_x,y}%
\end{equation}
\end{widetext}
where $\sigma_{\mathrm{P}}$, $\sigma_{\mathrm{H}}$, and $\sigma_{\parallel}$ are the local Pedersen, Hall,
and parallel conductivities, respectively \cite{Kelley:Ionosphere2009}.
Equation~(\ref{J_x,y}) is applicable to arbitrary $\hat
{\sigma}(z)$. For typical electrojet conditions, $\sigma_{\parallel}\gg
\sigma_{\mathrm{H}}\gg\sigma_{\mathrm{P}}$, this local solution
shows that $|\vec{J}|$ increases sharply
at $I\sim (\sigma_{\mathrm{P}}/\sigma_{\parallel})^{1/2}$, but in the vicinity of the magnetic equator ($I\rightarrow 0$) this
idealized 1D model predicts no electrojet, in contradiction to what
exists in nature.   Model~3 gives a reasonable approximation of an
equatorial electrojet but requires a non-constant $\vec B$.

We can now check whether Eq.~(\ref{J_x,y}) would predict electrojet
formation in accord with the above criterion of Eq.~(\ref{criterion}) or (\ref{generalized_sootno}).
For any arbitrary neutral gas density, Eq.~(\ref{generalized_sootno}) reduces to
$B_{z} (d\vec{U}/dz) \neq 0 $.  This requires a finite $B_{z}=B \sin I
\neq 0$, along with a vertical velocity shear,
$d\vec{U}/dz \neq 0$. This appears to create a contradiction because
Eq.~(\ref{J_x,y}) does not requires a velocity shear.
However, an understanding of boundary conditions resolves this
problem.  A shear must exist to fulfill the original assumption
$U_{x,y}(z)\rightarrow0$ as $z\rightarrow0,\infty$ but a non-zero $U$
within the system.

\begin{figure}
\includegraphics[width=0.47\textwidth]{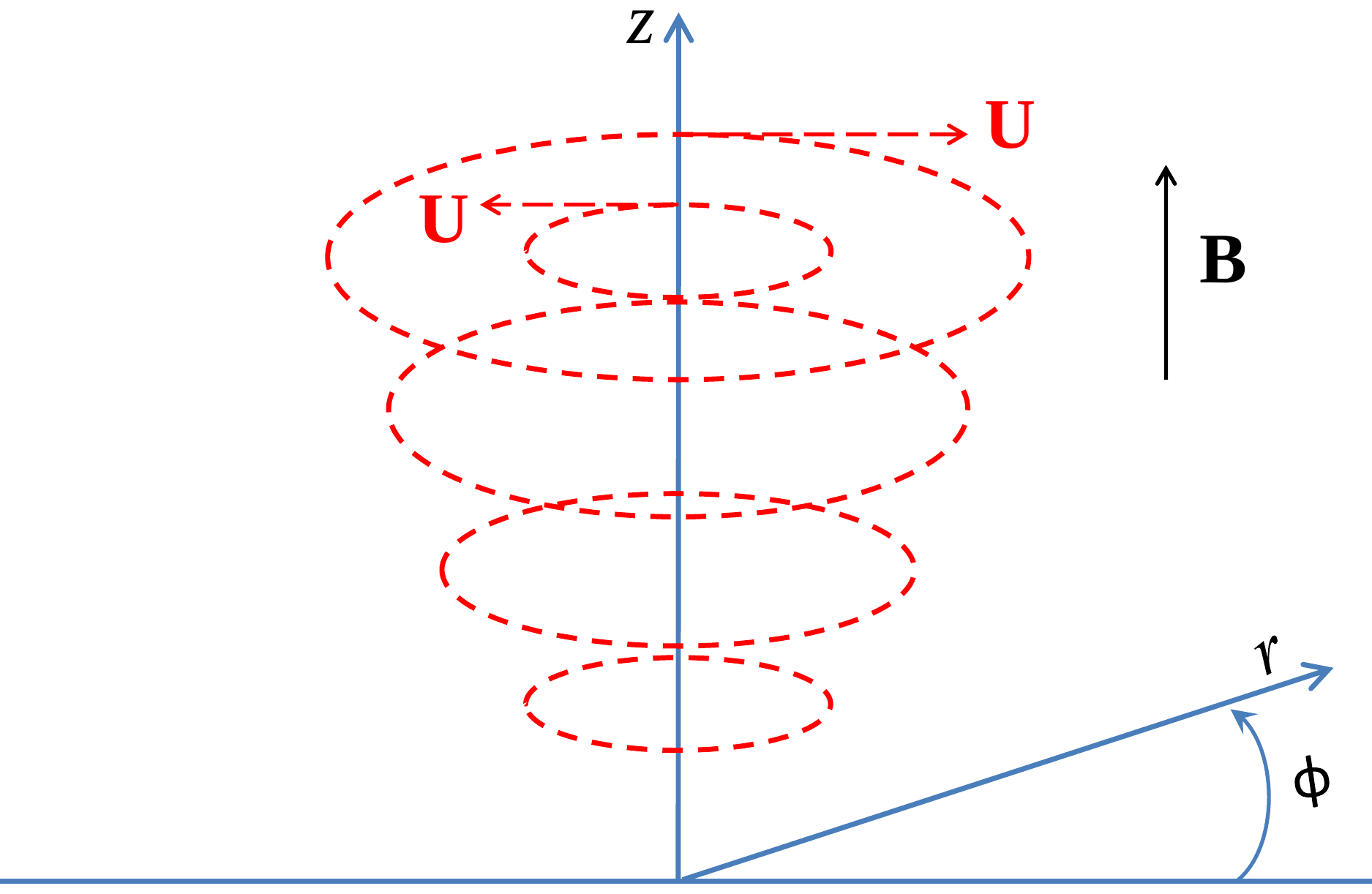}%
\caption{\label{Model_2_Cartoon} Axially symmetric flow of a neutral gas with possible differential rotation around a vertical uniform magnetic field, $\vec{B}$. The dashed red curves schematically show the circular flow lines with some velocities $\vec{U}(r,z)$ shown by vectors.}
\end{figure}

Model 2 is an axially symmetric 2-D system in cylindrical coordinates $r$,
$\phi$, $z$ (see Fig.~\ref{Model_2_Cartoon}) with a uniform vertical magnetic
field, $\vec{B}$, and a horizontally stratified neutral flow with differential
rotation, $U_{\phi}\left(r,z\right)  $.  For simplicity, we assume that the conductivity is only
height-dependent, $\hat{\sigma}=\hat{\sigma}(z)$ ($0<z<\infty$). This
problem has relevance to MHD generators \cite{Rosa:Plasmas91,Bokil_Toward15}.

In the axially symmetric geometry, dynamo Eq.~(\ref{initial}) becomes
\begin{equation}
\frac{1}{r}\ \frac{\partial}{\partial r}\left(  r[\hat{\sigma}(\vec{E}+\vec
{U}\times\vec{B})]_{r}\right)  +\frac{\partial J_{z}}{\partial z}%
=0.\label{rotated}%
\end{equation}
Assuming $\sigma_{\parallel} \gg \sigma_{\mathrm{H},\mathrm{P}}$, we have $ J_{z}\approx\sigma_{\parallel}E_{z}$ and
$|E_{z}| \ll |E_{r}|$, so that $\Phi(r,z)\approx\Phi_{0}(r)$.
Integrating (\ref{rotated}) over $z$, under assumption of $\left.  J_{z}\right\vert
_{z=0,\infty}=0$, we obtain the dominant radial electric field, $E_{r}%
(r)=-\left.  B\int_{0}^{\infty}\sigma_{\mathrm{P}}U_{\phi}dz\right/  \int%
_{0}^{\infty}\sigma_{\mathrm{P}}dz$, and the corresponding current density,%
\begin{eqnarray}
& & J_{r}(r,z) = \frac{B\sigma_{\mathrm{P}}(z)}{\int_{0}^{\infty}\sigma_{\mathrm{P}%
}(z^{\prime})dz^{\prime}}\times \nonumber\\
& &
\!\!\!\!\!\!\!\!\!\!\!\!\left(  U_{\phi}(r,z)\int_{0}^{\infty}\sigma
_{\mathrm{P}}(z^{\prime})dz^{\prime}-\int_{0}^{\infty}\sigma_{\mathrm{P}%
}(z^{\prime})U_{\phi}(r,z^{\prime})dz^{\prime}\right)\!  .\label{j_r}%
\end{eqnarray}
The axial current density is $J_{\phi}=(\sigma_{\mathrm{H}}/\sigma
_{\mathrm{P}})J_{r}$. In this approximation, the parallel current,
$J_{z}$, is found from the quasi-neutral charge conservation, $\partial
J_{r}/\partial r+\partial J_{z}/\partial z=0$, rather than from $J_{z}%
\approx\sigma_{\parallel}E_{z}$. Simple integration yields:
\begin{eqnarray}
& & J_{z} =\frac{B}{\int_{0}^{\infty}\sigma_{\mathrm{P}}(z^{\prime})dz^{\prime
}}\nonumber\\
&\times & \left[  \int_{0}^{z}\sigma_{\mathrm{P}}(z^{\prime\prime})dz^{\prime\prime
}\int_{0}^{\infty}\sigma_{\mathrm{P}}(z^{\prime})\ \frac{\partial U_{\phi
}(r,z^{\prime})}{\partial r}\ dz^{\prime}\right.  \nonumber\\
& - & \left.\ \int_{0}^{\infty}\sigma_{\mathrm{P}}(z^{\prime})dz^{\prime}%
\int_{0}^{z}\sigma_{\mathrm{P}}(z^{\prime\prime})\ \frac{\partial U_{\phi
}(r,z^{\prime\prime})}{\partial r}\ dz^{\prime\prime}\right] \! .\label{j_II}%
\end{eqnarray}

\begin{figure}
\includegraphics[width=0.47\textwidth]{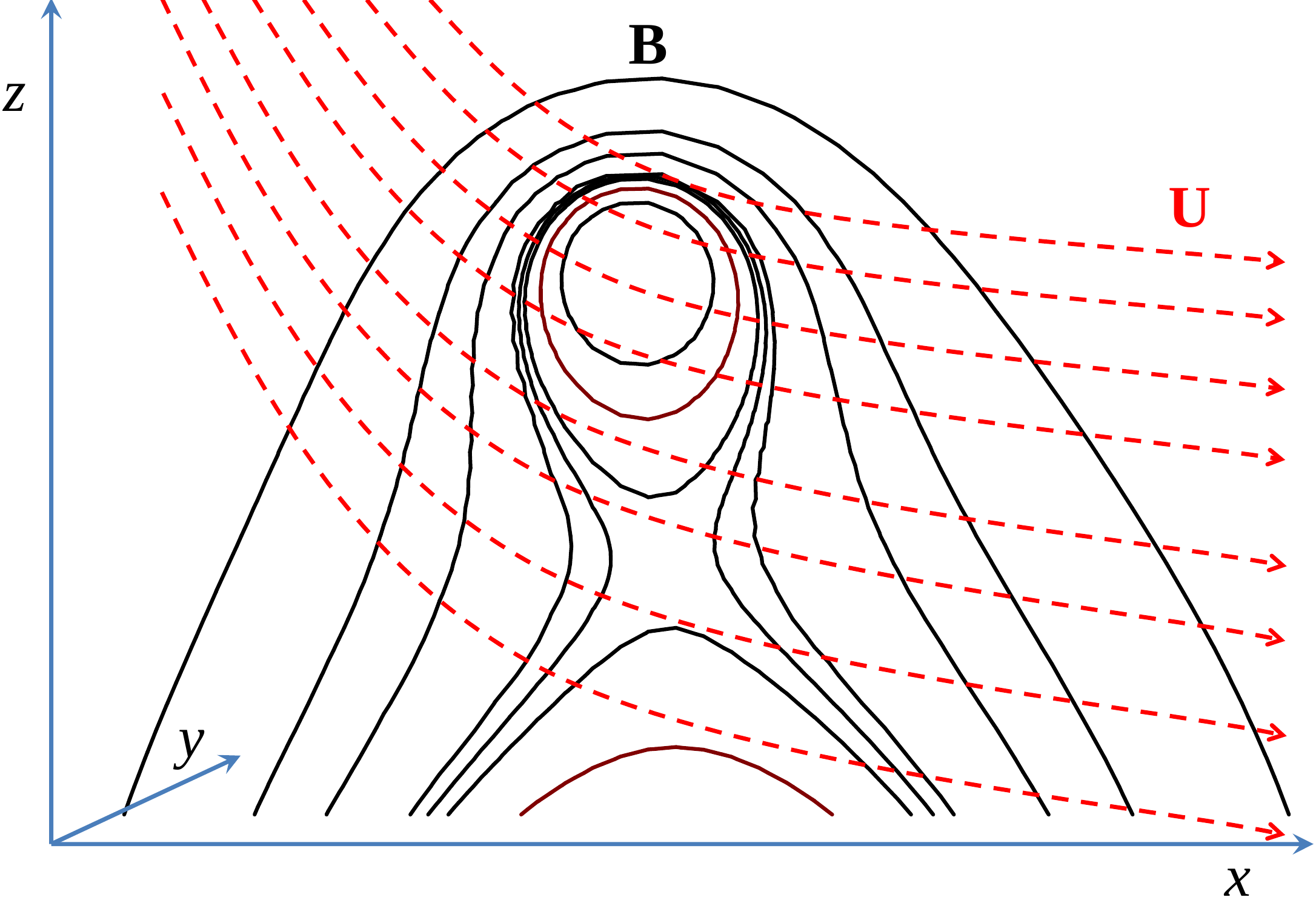}%
\caption{\label{Model_3_Cartoon}Cartoon of a more general gas flow and magnetic field in the 2D geometry. The solid black curves show the magnetic field lines in the $x,z$-plane, while the dashed red curves show the neutral gas flow lines (the flow velocity $\vec{U}$ may have also a $y$-component). The entire structure is invariant along $y$.
}
\end{figure}

The criterion for electrojet formation predicted by Eq.
(\ref{generalized_sootno}) requires just a vertical velocity shear, $\partial
U_{\phi}/\partial z \ne 0$. Equations~(\ref{j_r}) and (\ref{j_II}) imply
$U_{\phi}=U_{\phi}(r,z)$; otherwise $U_{\phi}(r)$ and $\partial U_{\phi
}/\partial r$ could be factored out from all integrals, leading to $\vec{J}%
=0$. This shear does not have to be present in the entire electrojet,
only in one or more locations where $\sigma_{\mathrm{P}}$ is not too small.

The last problem is similar to the first except it allows for
magnetic field lines arbitrarily distributed in
the $x,z$-plane, $\vec{B}=\left(  B_{x},0,B_{z}\right)  $, with $B_{x,z}$
invariant along the $y$-axis (see Fig.~\ref{Model_3_Cartoon}). In the main region of the weakly
ionized plasma, the magnetic field lines can be open or closed (assuming
intense localized currents in the $y$ direction). The neutral gas
velocity $\vec{U}$ may have all three spatial components, invariant along the
$y$-axis, $\vec{U}=\vec{U}(x,z)$. Depending on the specific realization, this
model can serve as a reasonable 2D approximation to both the equatorial electrojet and
Solar chromosphere.

This geometry suggests introducing curvilinear coordinates, $q_{i}$, where
$q_{1}$ specifies a given magnetic field line in the $x,z$-plane,
$q_{2}=y$, and $q_{3}$ specifies a location along the given field line. The $q_{2}$
coordinate line is orthogonal to the two others, but the $q_{1}$ and $q_{3}$
coordinate lines are not necessarily orthogonal to each other. We will
characterize any vector $\vec{S}$ either by its covariant ($S_{i}=\vec{S}%
\cdot\hat{\varepsilon}_{i}$) or contravariant ($S^{i}=\vec{S}\cdot
\hat{\varepsilon}^{i}$) components, $\vec{S}=S_{i}\hat{\varepsilon}^{i}%
=S^{i}\hat{\varepsilon}_{i}$, repeating
subscripts and superscripts implies summation. Here $\hat{\varepsilon}^{i}=\partial
q_{i}/\partial\vec{r}$ and $\hat{\varepsilon}_{i}=\partial\vec{r}/\partial
q_{i}$ are the basis vectors related via $\hat{\varepsilon}_{i}=g_{ik}%
\hat{\varepsilon}^{k}$, where $g_{ik}=\hat{\varepsilon}_{i}\cdot
\hat{\varepsilon}_{k}$ is the metric tensor.

In these curvilinear coordinates, Eq.~(\ref{initial}) reduces to%
\begin{equation}
\frac{\partial}{\partial q_{1}}(\sqrt{g}\,J^{1})+\frac{\partial}{\partial q_{3}%
}(\sqrt{g}\,J^{3})=0,\label{div_J_curvi}%
\end{equation}
where $J^{k}=[\hat{\sigma}(-\nabla\Phi+\vec{U}\times\vec{B})]^{k}$ and
$g\equiv\det\left(  g_{ij}\right)  =g_{11}g_{33}-g_{13}^{2}$.
As in the previous model, using $\sigma_{\parallel}\gg\sigma_{\mathrm{P},\mathrm{H}}$
we obtain $J^{3}\approx\sigma_{\parallel}{(-\nabla\Phi+\vec{U}\times\vec{B})}_{3}/g_{33}$ and $\Phi(q_{1},q_{3})\approx
\Phi_{0}(q_{1})$.
To obtain $\Phi_{0}(q_{1})$, we integrate Eq.~(\ref{div_J_curvi})
with respect to $q_{3}$, either over entire closed
field lines or over open field lines between sufficiently remote integration limits where $J^{3}\rightarrow0$. As a result, we obtain
\begin{equation}
\frac{\partial\Phi_{0}}{\partial q_{1}}\approx\frac{\int\left(\sigma_{P} \sqrt{\frac{g_{33}}{g}}\,K_{1}
 - \sigma_{H} K_{2}\right)
dl}{\int\sigma_{P}\sqrt{\frac{g_{33}}{g}}\,dl},\label{d_Psi_0}%
\end{equation}
\begin{widetext}
\begin{subequations}
\label{J^1,2}
\begin{eqnarray}
J^{1} & \approx & \frac{\sigma_{P}g_{33}\left(K_{1}\int\sigma_{P}
\sqrt{g_{33}/g}\,dl-\int\sigma_{P}K_{1}\sqrt{g_{33}/g}
\,dl+\int\sigma_{H}K_{2}dl\right)}{g\int\sigma_{P}\sqrt{g_{33}/g}\,dl}\,
-\,\sigma_{H}K_{2}\sqrt{\frac{g_{33}}{g}},\label{J^1}\\
J^{2} & \approx & \frac{\sigma_{H}\sqrt{g_{33}/g}\left(K_{1}\int\sigma
_{P}\sqrt{g_{33}/g}\,dl-\int\sigma_{P}K_{1}\sqrt{g_{33}/g
}\,dl+\int\sigma_{H}K_{2}dl\right)}{\int\sigma_{P}\sqrt{g_{33}/g}\,dl}\,+\, \sigma
_{P}K_{2},\label{J^2}
\end{eqnarray}
\end{subequations}
\end{widetext}
where $\vec{K}=\vec{U}\times\vec{B}$ and
$dl=\sqrt{g_{33}}\,dq_{3}$ is the length element along $\vec{B}$.
As in the previous model, $J^{3}$ can be obtained from Eq.~(\ref{div_J_curvi})
by simple integration.

Now we verify that $\nabla\times\vec{K}=0$ leads to $\vec{J}=0$. Indeed, in
this case $\vec{K}=\nabla\Psi$, so that $K_{1}=\partial\Psi/\partial q_{1}$
and $K_{2}=0$. This still leaves $J^{1,2}$ with two competing terms $K_{1}%
\int\sigma_{P}\sqrt{g_{33}/g}\ dl$ and $\int\sigma_{P}K_{1}\sqrt{g_{33}%
/g}\ dl$, which do not cancel each other if $K_{1}$ depends on $q_{3}$.
However, $\vec{U}\times\vec{B}=\nabla\Psi$ is orthogonal to $\vec{B}$, meaning
that magnetic field lines are `equipotential' with respect to $\Psi$ (as they
are with respect to $\Phi_{0}$). Then $K_{1}=\partial\Psi/\partial q_{1}$ can
be factored out from the corresponding integral, resulting in perfect
cancellation of the above terms and $J^{1,2}=0$. So, this
model also confirms the fact that irrotational $\vec{U}\times\vec{B}$ cannot
form electrojets, whatever the conductivities.


This paper provides a universal criterion for large-scale convective neutral
flows to form electrojets in weakly ionized plasmas. This criterion is
expressed in two equivalent forms by Eqs.~(\ref{irrotational}) and
(\ref{generalized_sootno}).
Although the vector field $\vec{U}\times\vec{B}$ is the neutral-dynamo
term, it is
$\nabla\times\vec{K}\equiv\nabla\times(\vec{U}\times\vec{B})-\partial\vec{B}/\partial t$
that determines the actual dynamo, as expressed by
Eq.~(\ref{urka}). This driver plays for generation of the electrojet
the same role as the DC electric current plays for generation of the
magnetic field (the Biot-Savart law). The above criterion should be taken into account 
when modeling the neutral dynamo in space or laboratory plasmas.

This work was supported by NSF/DOE Grant PHY-1500439, NASA Grants NNX11A096G and NNX14AI13G, and NSF-AGS Postdoctoral Research Fellowship Award No.~1433536.

%

\end{document}